\documentclass[preprint,12pt]{elsarticle}

\usepackage{amssymb}
\usepackage{subfigure}
\usepackage{booktabs}
\usepackage{tabularx}
\usepackage{array}
\usepackage{lscape}
\usepackage{longtable}
\usepackage{epstopdf}
\usepackage{graphicx}    
\usepackage{booktabs}    
\usepackage{colortbl}    
\usepackage{threeparttable}
\usepackage{rotating}

\newtheorem{definition}{Definition}[section]

\newcommand{\PreserveBackslash}[1]{\let\temp=\\#1\let\\=\temp}
\newcolumntype{C}[1]{>{\PreserveBackslash\centering}p{#1}}
\newcolumntype{R}[1]{>{\PreserveBackslash\raggedleft}p{#1}}
\newcolumntype{L}[1]{>{\PreserveBackslash\raggedright}p{#1}}

\journal{Physica A} 

\begin{document}

\begin{frontmatter}



\title{Modelling the Self-similarity in Complex Networks Based on Coulomb's Law}


\author[address1]{Haixin Zhang}
\author[address1,address2]{Daijun Wei}
\author[address3]{Yong Hu}
\author[address1]{Xin Lan}
\author[address1,address4]{Yong Deng\corref{label1}}
\cortext[label1]{Corresponding author: Yong Deng, School of
Computer and Information Science, Southwest University,
Chongqing, 400715, China. Email address: ydeng@swu.edu.cn;
professordeng@163.com}

\address[address1]{School of Computer and Information Science, Southwest University, Chongqing 400715, China}
\address[address2]{School of Science, Hubei University for Nationalities, Enshi 445000, China}
\address[address3]{Institute of Business Intelligence and Knowledge Discovery, Guangdong University of Foreign Studies, Guangzhou, 510006, China}
\address[address4]{School of Engineering, Vanderbilt University, Nashville, TN, 37235, USA}

\begin{abstract}
Recently, self-similarity of complex networks have attracted much attention. Fractal dimension of complex network is an open issue. Hub repulsion plays an important role in fractal topologies. This paper models the repulsion among the nodes in the complex networks in calculation of the fractal dimension of the networks. The Coulomb's law is adopted to represent the repulse between two nodes of the network quantitatively. A new method to calculate the fractal dimension of complex networks is proposed. The Sierpinski triangle network and some real complex networks are investigated. The results are illustrated to show that the new model of self-similarity of complex networks is reasonable and efficient.
\end{abstract}


\end{frontmatter}


\section{Introduction}
Network science plays a more and more important role in academic researches\cite{barabasi2004network,furuya2008generalized,serrano2009extracting,gao2011networks,barthelemy2011spatial,pu2012robustness}. Recently, complex networks have attracted much attention in diverse areas of science and technology\cite{strogatz2001exploring,dorogovtsev2008critical,shao2009structure,Gao20135490,gao2013bio,Wei20132564,du2014new}. Representations of complex systems by complex networks has proven to be generally successful to describe their various features\cite{da2006hierarchical,costa2008concentric,silva2010identifying,silva2012local,yu2013visibility}. It has been shown that the small-world\cite{watts1998collective} property and the scale-free\cite{barabasi1999emergence} property are the two fundamental properties of complex networks. The small-world property means that the average shortest path length between nodes in the network increase very slowly while the number of nodes increase largely. The scale-free property means that a small number of nodes of the network have a large number of neighbor nodes, while the other large number of nodes of the network have a small number of neighbor nodes.

Fractal theory is applied to model complex systems\cite{bunde1994fractals,jurgens1992chaos}. The strict self-similarity property or the statistical self-similarity property is the typical characteristic of the fractal objects. Fractal dimension is regarded as an intrinsic characteristic used to describe the complexity of a fractal object\cite{bunde1994fractals}. Fractal and self-similarity have been found widely exist in nature\cite{furuya2011multifractality,carletti2010weighted,rozenfeld2009fractality,zou2007topological,zhou2007exploring,fan2012fractal}.
Song \emph{et al.} found that a variety of real complex networks consist of self-repeating patterns on all length scales in Ref.\cite{song2005self}. Recently, fractal and self-similarity of complex networks have attracted much attention\cite{radicchi2008complex,serrano2008self,rozenfeld2010small,carletti2010weighted,kawasaki2010reciprocal,amancio2011using,yakubo2011scale,dan2012multifractal}. Many researchers have analyzed the fractal property of complex networks and proposed different algorithms to calculate the fractal dimension of complex networks \cite{kim2007fractality,kim2007fractality2,blagus2012self,gallos2007review,moon2011core,zhang2013self,silva2012local}. One of the progress in this field is that Wei \emph{et al.} studied the self-similarity of weighted complex networks\cite{Wei20132564}.


In real application, the accurate value of the fractal dimension defined by Hausdorff\cite{bunde1994fractals} can not be obtained. Box-covering algorithm is a commonly used tool to calculate the fractal dimension, namely box dimension. Box-covering algorithm is to use the minimal number of boxes to cover the fractal objects. The number of boxes, denoted by $N(\varepsilon)$, and the size of the box, denoted by $\varepsilon$, yield the relationship: $
N\left( \varepsilon  \right) \sim \varepsilon ^{D_B }$, $ D_B $ is called the box dimension, it can be obtained by regressing $ \log \left( {N\left( \varepsilon  \right)} \right) $ vs. $\log \left( \varepsilon  \right)$. Box-covering algorithm is widely used to calculate the fractal dimensions of complex networks\cite{song2007calculate,schneider2012box,kim2007box,gao2008accuracy,daqing2011dimension} .


According to Song's research \cite{song2006origins}, during the self-similar dynamical evolution of complex networks, the hubs grow by preferentially linking with less-connected nodes to generate a more robust fractal topology. In other words, the nodes with high degree do not connect directly and dislike to being in the same box. It seems that a higher degree of hub repulsion plays a more important role in fractal topologies than non-fractals. As a result, it is necessary and reasonable to model the repulsion among the nodes in the complex networks in calculation of the fractal dimension of the networks, which is our motivation of this paper. In Coulomb's law\cite{baigrie2007electricity}, if the two charges have the same sign, the electrostatic force between them is repulsive. Inspired by the modeling of repulsion in Coulomb's law, we adopt this law to represent the repulse between two nodes quantitatively. In our proposed model, the connected nodes in the complex networks are regarded as "electric charges" and there exists "electrostatic interaction force" between them over the edge. The value of the repulsive force of each edge relies on the degrees of the two nodes linked by this edge. The connected nodes with higher degree have greater force over their link and thus they will be less likely to be covered by the same box in the box-covering process.

The rest of this paper is organized as follows. In section 2, some basic concepts are introduced. The new model of complex network inspired by Coulomb's law and the proposed method to obtain the fractal dimension are proposed in section 3. Some applications are illustrated in section 4. Finally, the conclusions are drawn in the last section.

\section{Preliminaries}
In this section, some basic concepts of complex network, the typical box-covering algorithm for fractal dimension of complex network and the Coulomb's law are introduced.

For a complex network $G={(N, E)}$ and $N={(1,2,\cdots,n)}$, $E={(1,2,\cdots,m)}$, where $n$ is the total number of the nodes, $m$ is the total number of the edges, and the cell $e_{ij}(i,j=1,2,\cdots,n)$ of the edges equals 1 if node $i$ is connected to node $j$, and 0 otherwise.

Degree is a basic indicator\cite{jamakovic2008relationships}, the degree of node $i$, denoted by $k_i$, equals the number of nodes which are connected to node $i$ directly. Degree describes the number of neighbors a node has. Shortest path between node $i$ and node $j$, denoted by $S_{ij}$, is a set of edges which can connect node $i$ and node $j$ and satisfy the number of such edges are minimized\cite{newman2004analysis}. Shortest path length between node $i$ and node $j$, denoted by $l_{ij}$, is the minimized number of such edges which connect node $i$ and node $j$. The diameter of a network, denoted by $d$, is the longest shortest path length among any two nodes.

The original definition of box-covering is initially proposed by Hausdorff\cite{jurgens1992chaos}. It is initially applied in the complex networks by Song, \emph{et al.}\cite{song2005self,song2007calculate}. For a given box size $l_{B}$, a box is a set of nodes where all shortest path length $l_{ij}$ between any two nodes $i$ and $j$ in the box are smaller than $l_B$. The minimum number of boxes used to cover the entire network, denote by $N_B$,  yield the following relationship:
\begin{equation}\label{fractal}
N_B  \sim {l_B} ^{ - D_B }
\end{equation}
$ D_B $ is the box dimension of complex network $G$, it can be obtained by regressing $ \log \left( {N_B } \right)$ vs. $\log \left( {{l_B } } \right)$.

\begin{figure}\begin{center}
\includegraphics[width=9cm,height=6cm]{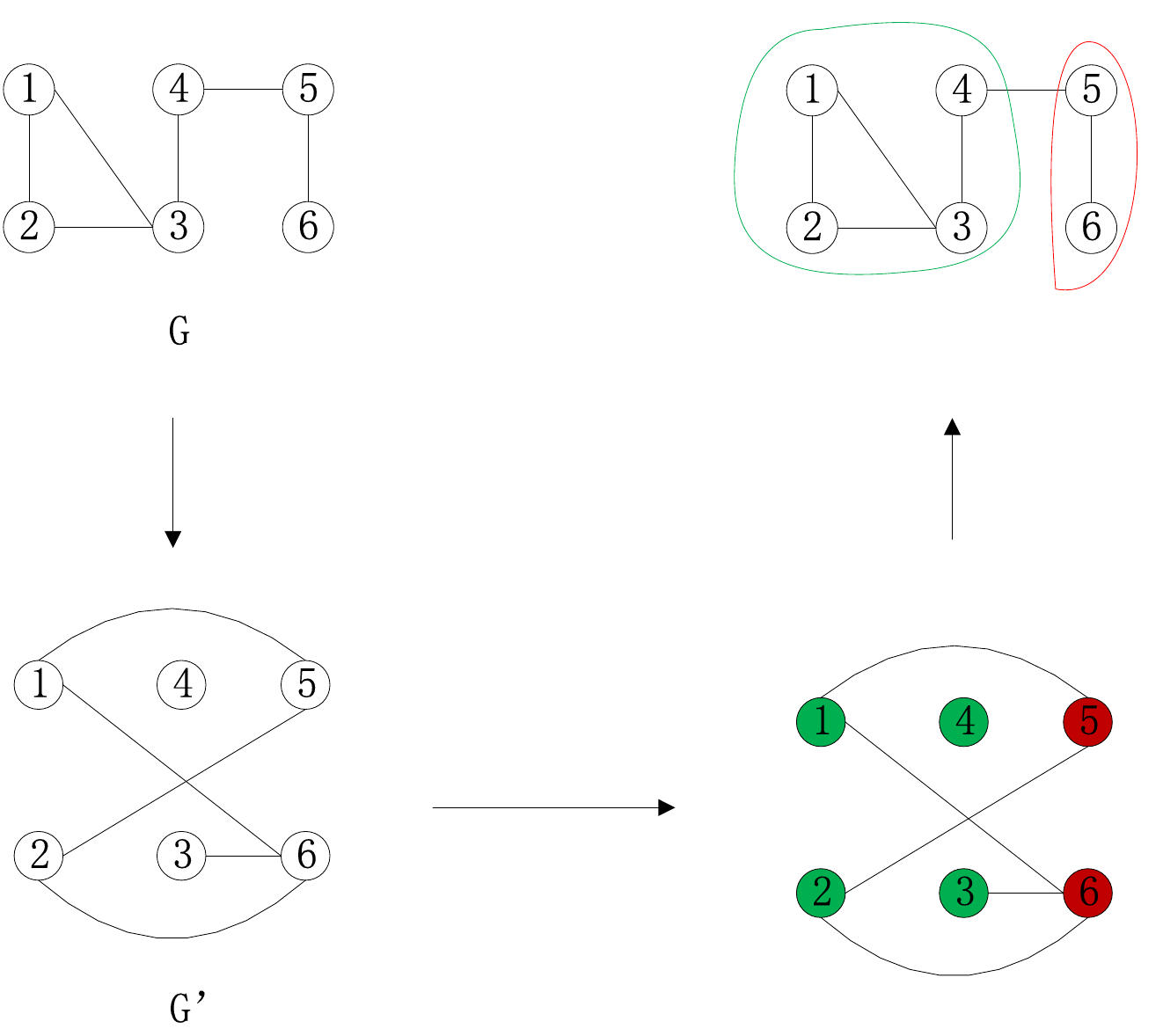}
\caption{An example of Song's method. Starting from $G_1$, a dual network $G_2$ is constructed for a given box size (here $l_B=3$),where two nodes are connected if their shortest path length $l_{ij} \ge l_{B}$. Then a greedy coloring algorithm is applied to determine the box covering in $G_2$, as shown in the plot.}
\label{previous}       
\end{center}
\end{figure}

In Song's box-covering algorithm, the length of every edge equals to 1 and the shortest path length between any two nodes is the minimum number of edges which connect them. The maximum shortest path length between any two nodes in the box cannot be greater than the box size. The fractal dimension is calculated via a greedy coloring algorithm. There are several main steps of Song's method showing as follows. An example of Song's method\cite{song2007calculate} is illustrated in Figure \ref{previous}.

\begin{description}
  \item[Transforming] For given network $G_{1}$ and  box size $l_{B}$. A new network $G_{2}$ is obtained, in which node $i$ is connected to node $j$ when $l_{ij} \ge l_{B}$.
  \item[Greedy coloring] Mark as much as possible nodes in network $G_{2}$ with the same color, but the color of every node must be different from the colors of its nearest neighbors. A coloring network $G_{3}$ can be obtained.
  \item[Box counting] The nodes in the same color are in the same box, so the number of boxes equals the number of colors, the value of $N_{B}$ is obtained.
  \item[Regressing] Change the box size $l_{B}$, $ D_B $ can be obtained by regressing $ \log \left( {N_B } \right)$ vs. $\log \left( {{\rm{l}}_{\rm{B}} } \right)$.
\end{description}

This paper adopts Coulomb's law to represent the repulse between two nodes, so the Coulomb's law is briefly reviewed. Suppose $e_1$ and $e_2$ are two point charges, with electric charges $q_1$ and $q_2$ respectively and their distance $r$, then the electrostatic interaction force between them, denoted by $F$ could be calculated by\cite{baigrie2007electricity}
\begin{equation}\label{}
F = k_e \frac{{\left| {q_1 q_2 } \right|}}{{r^2 }}
\end{equation}
If the two charges have the same sign, the electrostatic force between them is repulsive; if they have different sign, the force between them is attractive. The Coulomb's law states that the magnitude of the Electrostatics force of interaction between two point charges is directly proportional to the scalar multiplication of the magnitudes of charges and inversely proportional to the square of the distances between them.

\section{Proposed method}
For a complex network, $G={(N, E)}$ and $N={(1,2,\cdots,n)}$, $E={(1,2,\cdots,m)}$, the edge value $e_{ij}(i,j=1,2,\cdots,n)$ equals 1 when node $i$ is connected to node $j$, and 0 otherwise. Inspired by Coulomb's law, the degree of each node is regarded as the "electric charge" of that node, and the "electric charge" of every node are with the same sign. The directly connected "electric charges" have repulsive force over the links. The edge repulsive force of the complex network $G$ is defined as follows:

\begin{definition}[edge repulsive force]
 Suppose $e_{ij}=1$, which is the edge which connects node $i$ and node $j$, let $k_i$ and $k_j$ be the degree of node $i$ and node $j$ respectively, and then the repulsive force of the edge $e_{ij}$, denoted by $f_{ij}$, could be calculated by
\begin{equation}\label{edgerepulsionforce}
f_{ij}  = k_i  \times k_j
\end{equation}
\end{definition}

It is noticed that the repulsive force only exists on the edges which connect the nodes directly. If $e_{ij}=0$, there is no edge connects node $i$ and node $j$ directly, then there is no direct repulsive force between node $i$ and node $j$. The value of the repulsive force of each edge depends on the both degrees of the two nodes linked by this edge. The connected nodes with higher degree have greater force over their link and thus they will be less likely to be covered by the same box in the box-covering process, which reflect the hub repulsion.

Similar to the shortest path and diameter in the traditional complex network, the smallest repulsive force path between any two nodes and the repulsive force diameter in our model is defined.

\begin{definition}[smallest repulsive force path]
Denote $fsp_{ij}$ as the smallest repulsive force path between node $i$ and node $j$, which satisfies
\begin{equation}\label{repulsionforcepath}
fsp_{ij}  = \min \left( {f_{it}  +  \ldots  + f_{tj} } \right)
\end{equation}
where $f_{it}$ and $f_{tj}$ are the edge repulsive force of edge $e_{it}$ and edge $e_{tj}$ respectively.
\end{definition}

\begin{definition}[repulsive force diameter]
Let $fsp_{ij}$ be the smallest repulsion force path between node $i$ and node $j$ of network $G={(N, E)}$. Then the repulsive force diameter of $G$, denoted by $fd$, is defined as
\begin{equation}\label{repulsionforcediameter}
fd = \max \left( {fsp_{ij} } \right)
\end{equation}
\end{definition}

\begin{figure}
\begin{center}
\includegraphics[width=9cm,height=6cm]{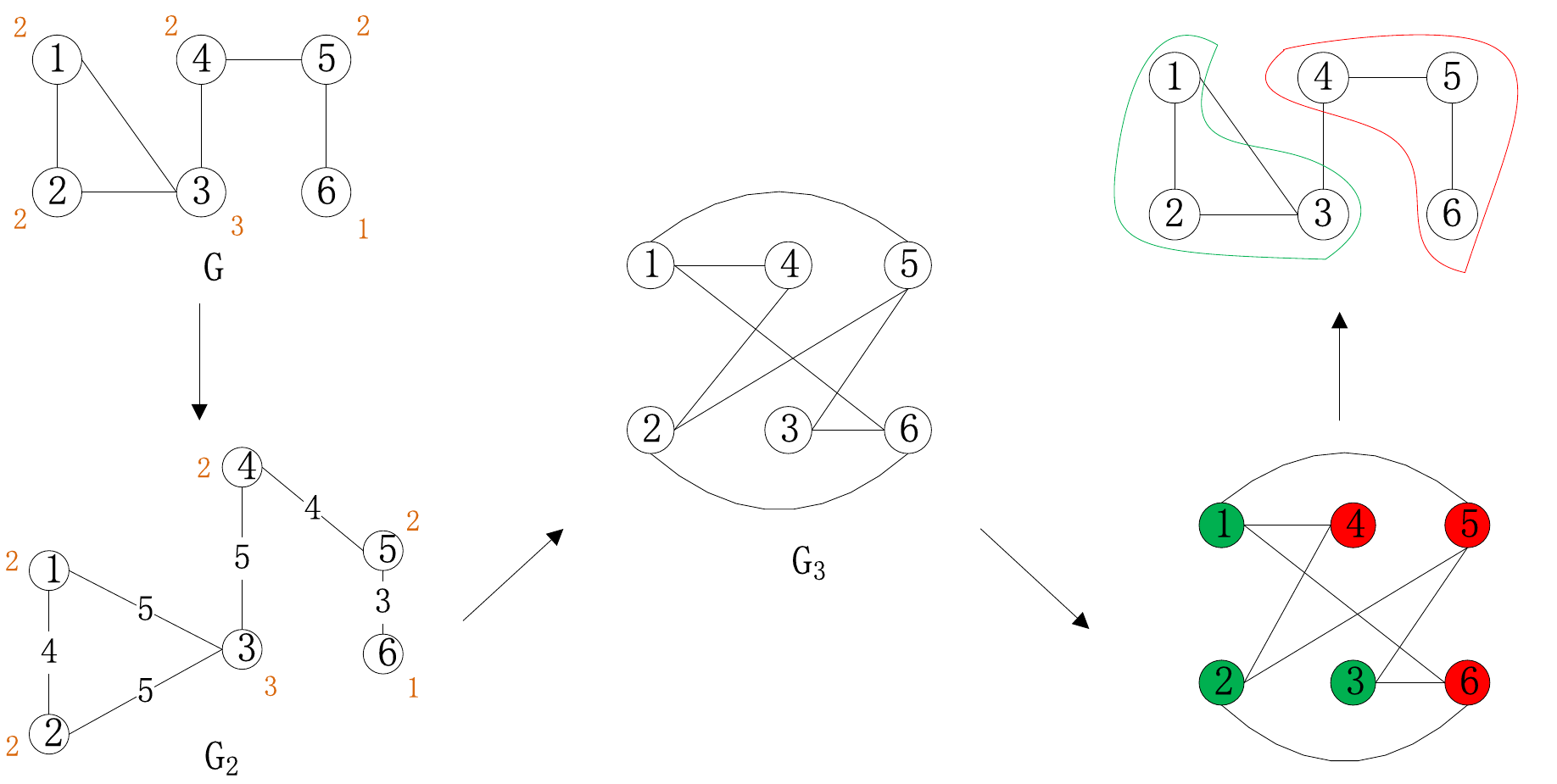}
\caption{An example of the proposed method. Starting from $G$, we construct a new network with edge repulsive force, $G_2$, by assigning the value of repulsive force of each edge based on the node degrees(marked around the node). Then a dual network $G_3$ is obtained for a given box size(here $l_{B}=10$), where two nodes are connected if their smallest repulsive force path $srfp \ge l_{B}$. Then a greedy coloring algorithm is applied to determine the box covering in $G_3$, as shown in the plot}
\label{proposed}
\end{center}
\end{figure}

Then the fractal dimension of $G$ will be obtained through the following several steps. An example of the proposed method is illustrated in Figure \ref{proposed}.

\begin{description}
  \item[step 1] Transform the network,$G$, into a new network with edge repulsion force,$G_2$.
  \begin{itemize}
    \item Calculate the degree of each node of the network.
    \item Calculate the repulsive force of each edge of network $G$, the value of the edge repulsion force could be calculated by Eq. (\ref{edgerepulsionforce}).
  \end{itemize}
  \item[step 2] Produce a dual network. A new network $G_3$ is obtained, in which node $i$ is connected to node $j$ when $fsp_{ij} \ge l_{B}$. $l_{B}$ is the box size, in which the smallest repulsion force path of any two nodes in this box is smaller than $l_{B}$.
  \item[step 3] Greedy coloring process. Color the nodes of the network $G_3$ using as minimal number of colors as possible, which satisfies that the color of one node is different from the colors of its neighbors.
  \item[step 4] Count the number of boxes $N_{B}$. Let one color represent one box in network $G_3$. Then the number of the used colors is the number of boxes needed to cover the network, denoted by $N_{B}$.
  \item[step 5] Change the box size $l_B$ and repeat the above steps 2-4. $l_B$ could be changed from the smallest repulsive force path to the next smallest repulsion force path till $fd$,the repulsion force diameter of network $G_2$.
  \item[step 6] Regress $\log \left( {N_B } \right)$ vs. $\log \left( {l_B } \right)$. $N_{B}$ and $l_B$ satisfy the relation:
        \begin{equation}\label{relation}
        N_B  \sim l_B ^{ - D_F }
        \end{equation}
    The slope of the regressed line is the negative value of the fractal dimension of $G$, denoted by $D_F$.
\end{description}


In the example of the proposed method, the repulsive force diameter of network $G_2$ is 18(1-3-4-5-6 or 2-3-4-5-6), the box size $l_B$ could be changed from the smallest repulsive force path 2(5-6) to 4(1-2 or 4-5) and then 6(1-3 or 2-3 or 3-4 or 4-5-6) and then 10(3-4-5) and then 12(3-4-5-6) and then 18. Using the box-covering algorithm, we can calculate the corresponding number of boxes $N_B$ to cover the network vs the given $l_B$. Then the fractal dimension could be obtained by regressing $\log \left( {N_B } \right)$ vs. $\log \left( {l_B } \right)$.

In our model, the repulsive force is redefined as the metric in complex networks. The metric in complex network is not the traditional shortest path length but the redefined edge repulsive force. The previous method is only from the view of the topology structure of the network, while our proposed method is not only from the view of the topology structure but also from the view of the hub repulsion of the network. Under this redefinition, a similar greedy coloring algorithm is deployed to calculate the fractal dimension. In the next section, some complex networks are investigated and the results show that the minimum number of the boxes used to cover the networks and the box size follows a power law rule, which reflects the self-similarity of the complex networks.

\section{Applications and discussions}
\begin{figure*}
\begin{center}
\includegraphics[width=0.9\textwidth]{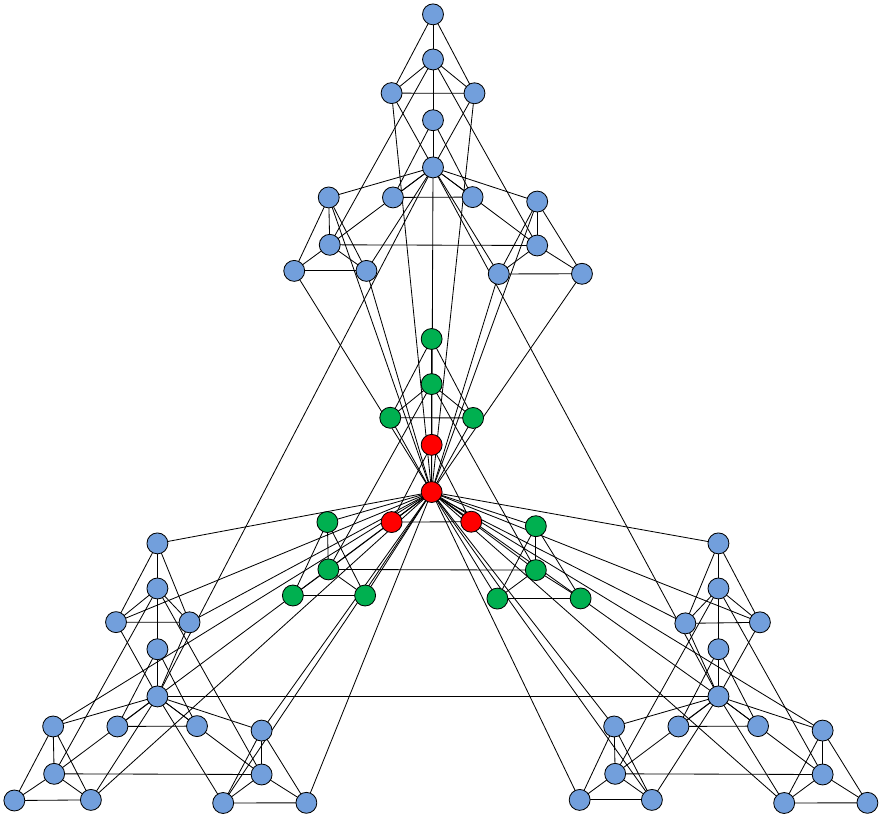}
\caption{Sierpinski triangle network: The starting point of this construction is a small cluster of four densely linked nodes(see the four central nodes). Next, three replicas of this module are generated. The three external nodes of the replicated clusters connected to the central node of the old cluster and the central node of the replicated clusters connected to each other. These 16 nodes produce a large 16-node module. Three replicas of this 16-node module are then generated and the peripheral nodes of the replicated module connected to the central node of the old module and the central node of the replicated module connected to each other, which produces a new module of 64 nodes.}
\label{Sierpinski}
\end{center}
\end{figure*}

\begin{figure}
\centering
\begin{minipage}{0.9\textwidth}
\subfigure{\includegraphics[width=0.9\textwidth]{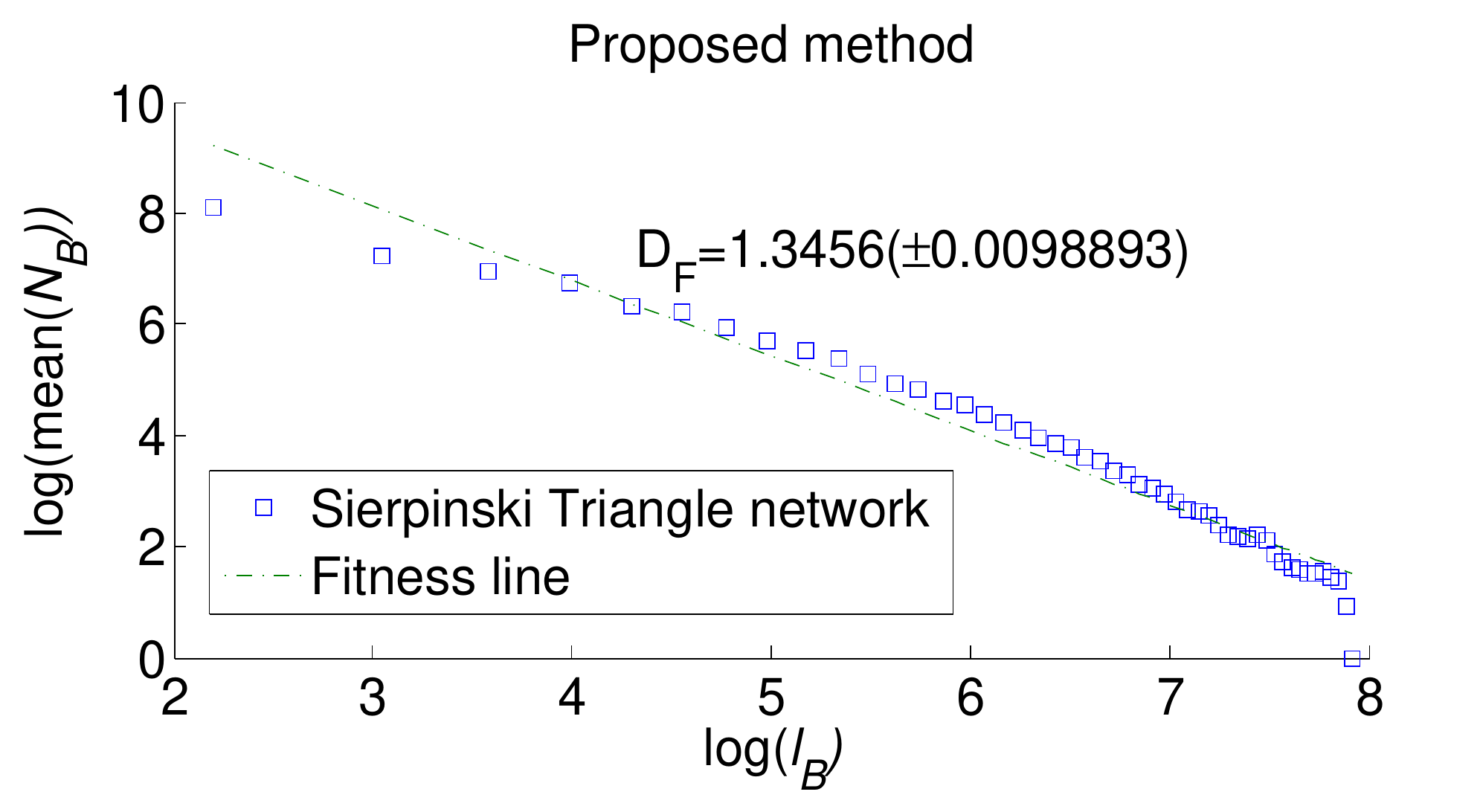}} \\
\subfigure{\includegraphics[width=0.9\textwidth]{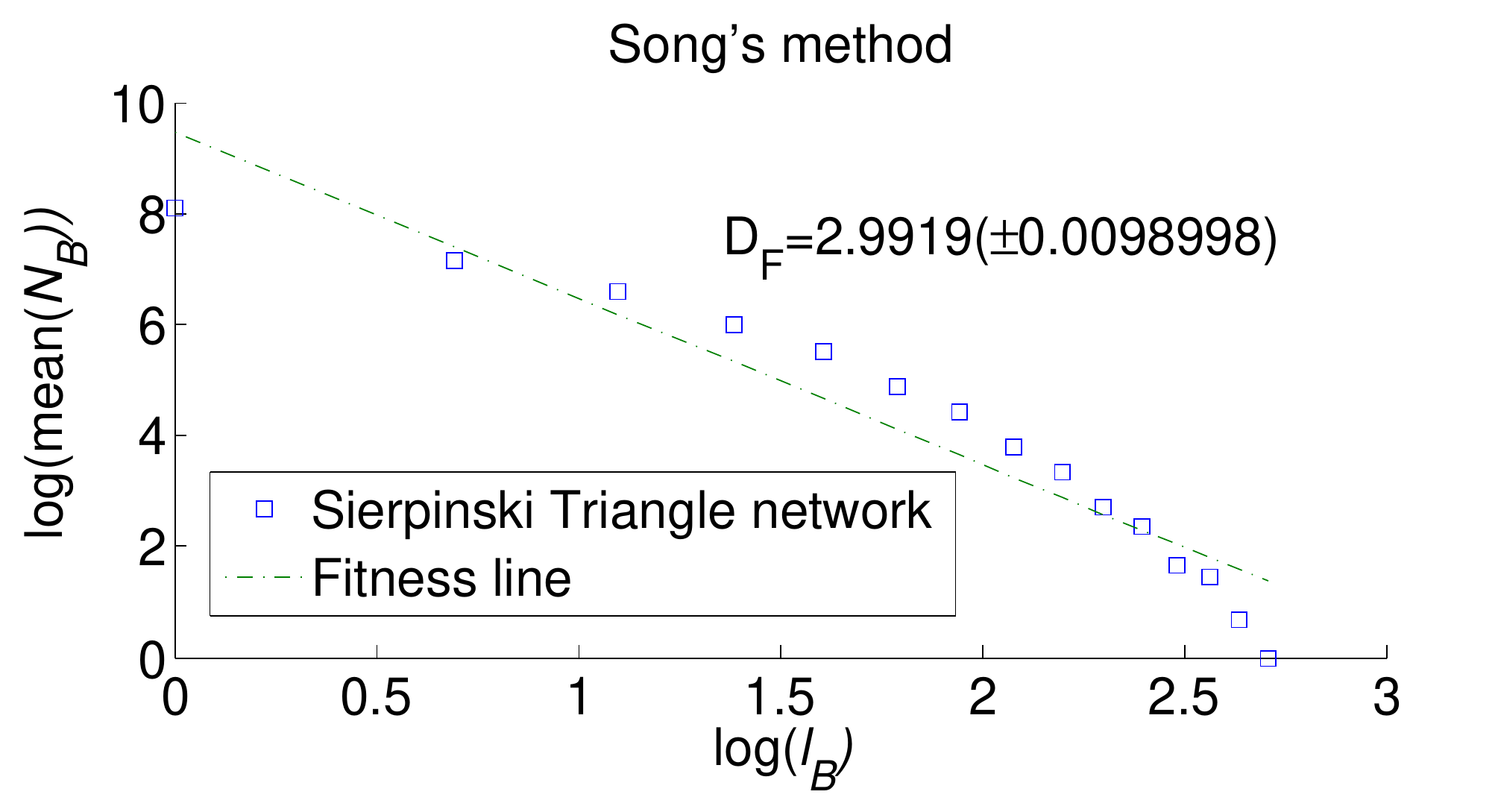}}
\end{minipage}
\caption{The $N_{B}$ vs. $l_B$ of Sierpinski triangle network obtained in a log-log scale. The vertical ordinate of every subplot is the mean value of $N_{B}$ for 1000 times, and the horizontal ordinate represents the box size, $l_B$. Regressing $\log \left( {N_B } \right)$ vs. $\log \left( {l_B } \right)$ for different values of $l_B$, we obtained the fractal dimensions. The mean fractal dimension with the standard deviations in parentheses for 1000 times are marked in every subplot both by our proposed method and Song's method.\label{results1}}
\end{figure}

The Sierpinski triangle is a classic fractal structure, whose theoretical value of fractal dimension equals to $\log \left( 3 \right)/\log \left( 2 \right)$($ \approx 1.5850$). The Sierpinski triangle network\cite{barabasi2004network} is produced by Self-replicating(see Figure \ref{Sierpinski}). We applied our proposed method and Song's method to the Sierpinski triangle network. The node coloring process is an NP-hard problem\cite{Garey:1990:CIG:574848}, the results of the greedy coloring algorithm may depend on the original coloring sequence. In order to investigate the quality of the algorithm, we randomly reshuffle the coloring sequence and apply the greedy algorithm for 1000 times in both methods. A Sierpinski triangle network of 3282 nodes is studied. The results are shown in Figure \ref{results1}. It can be seen that the number of the boxes used to cover the Sierpinski triangle network follows a power law with respect to the box size both by our proposed method and Song's method. The small standard deviations indicates that the greedy coloring algorithm works very well both in our proposed method and Song's method. The power law shows that both methods are efficient to discover the self-similarity property of the Sierpinski triangle network. However, the calculated value of the fractal dimension by our proposed method(1.3456) is closer to the theoretical value of fractal dimension of the Sierpinski triangle(1.5850) than that of Song's method(2.9919). It is proved that the modeling of hub repulsion is reasonable and necessary to capture the self-similarity of complex network.

\begin{figure}
  \centering
\subfigure{
    \label{fig:subfig:a} 
    \includegraphics[width=0.45\textwidth]{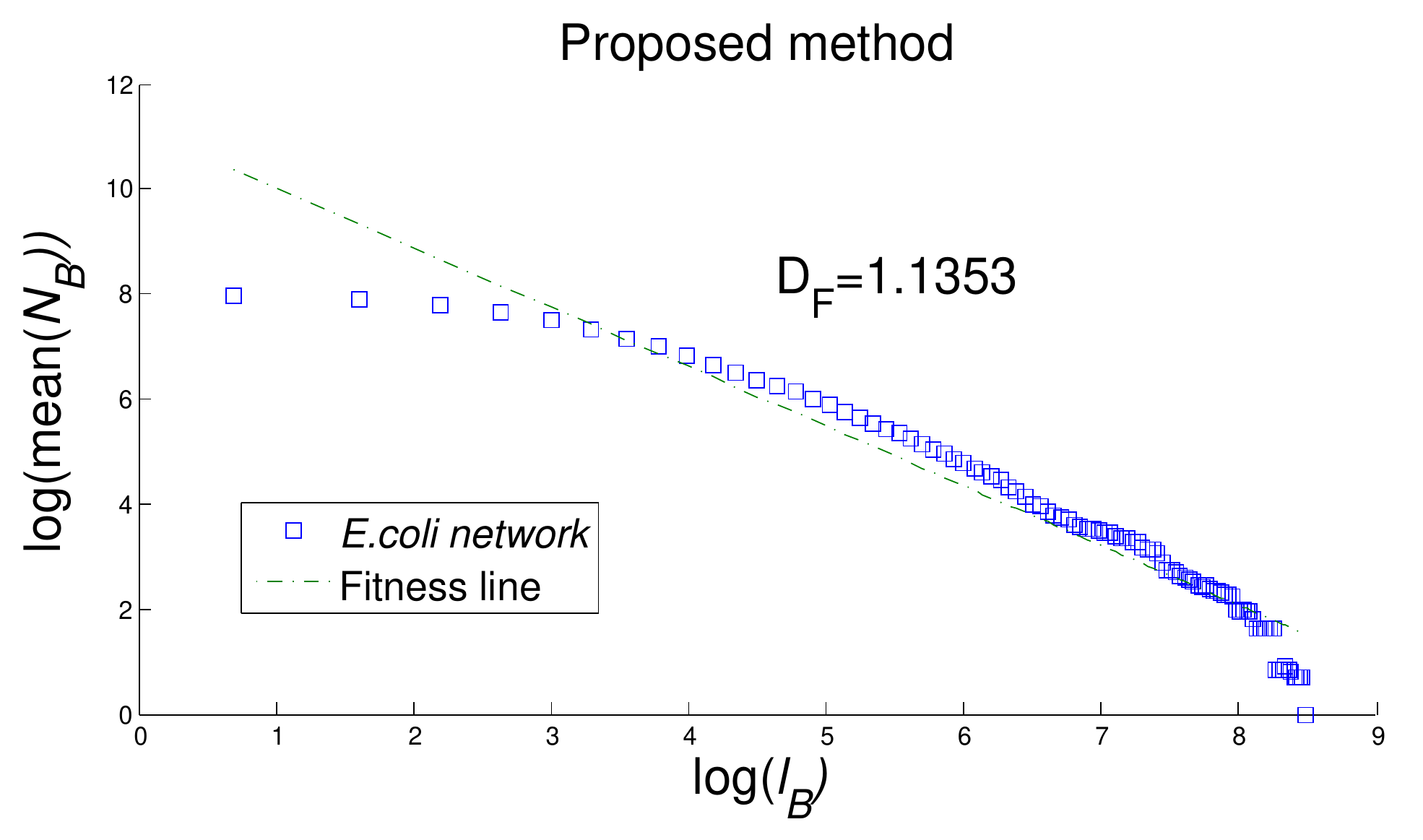}}
  \subfigure{
    \label{fig:subfig:a'} 
    \includegraphics[width=0.45\textwidth]{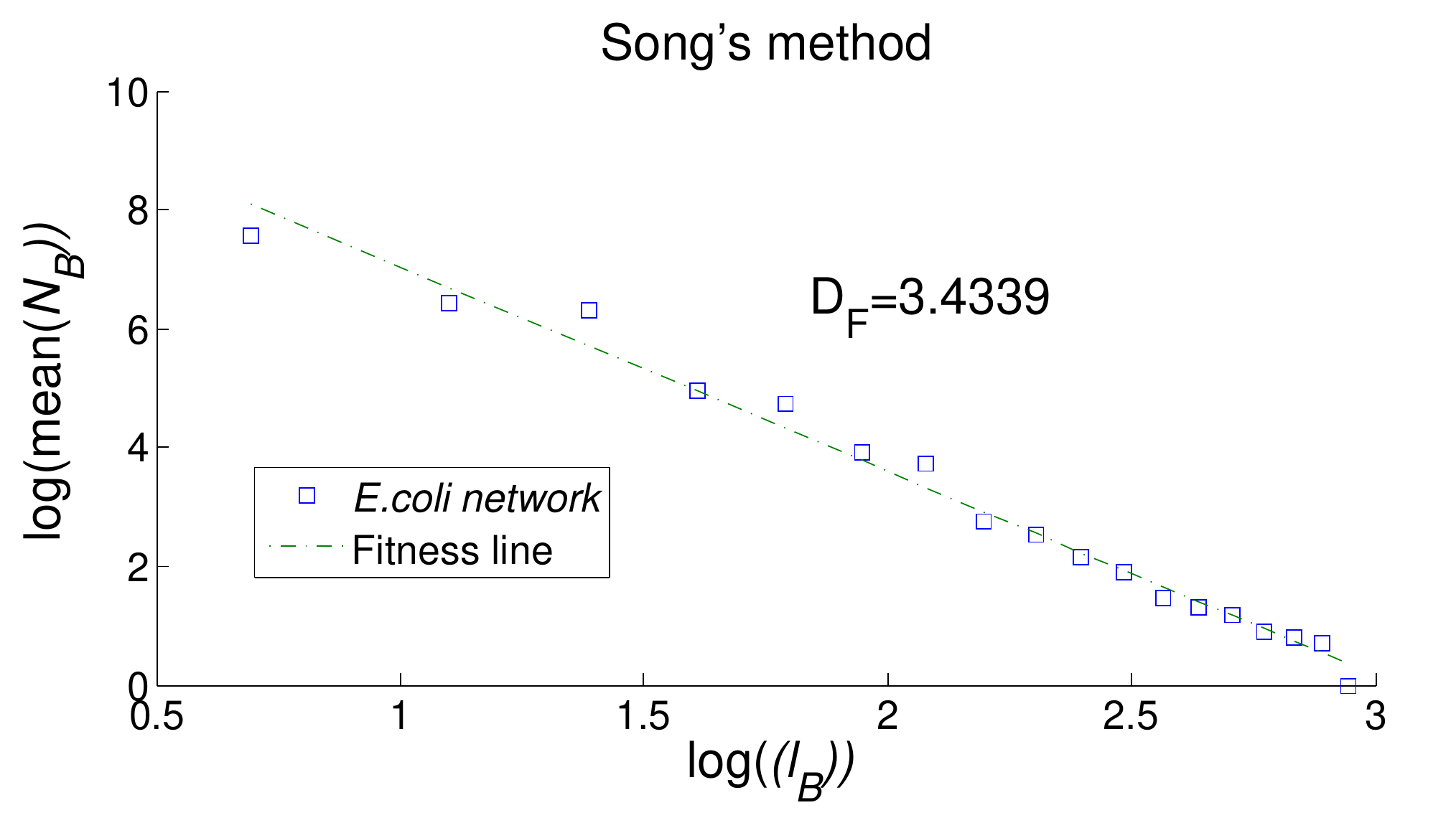}}
  \subfigure{
    \label{fig:subfig:b} 
    \includegraphics[width=0.45\textwidth]{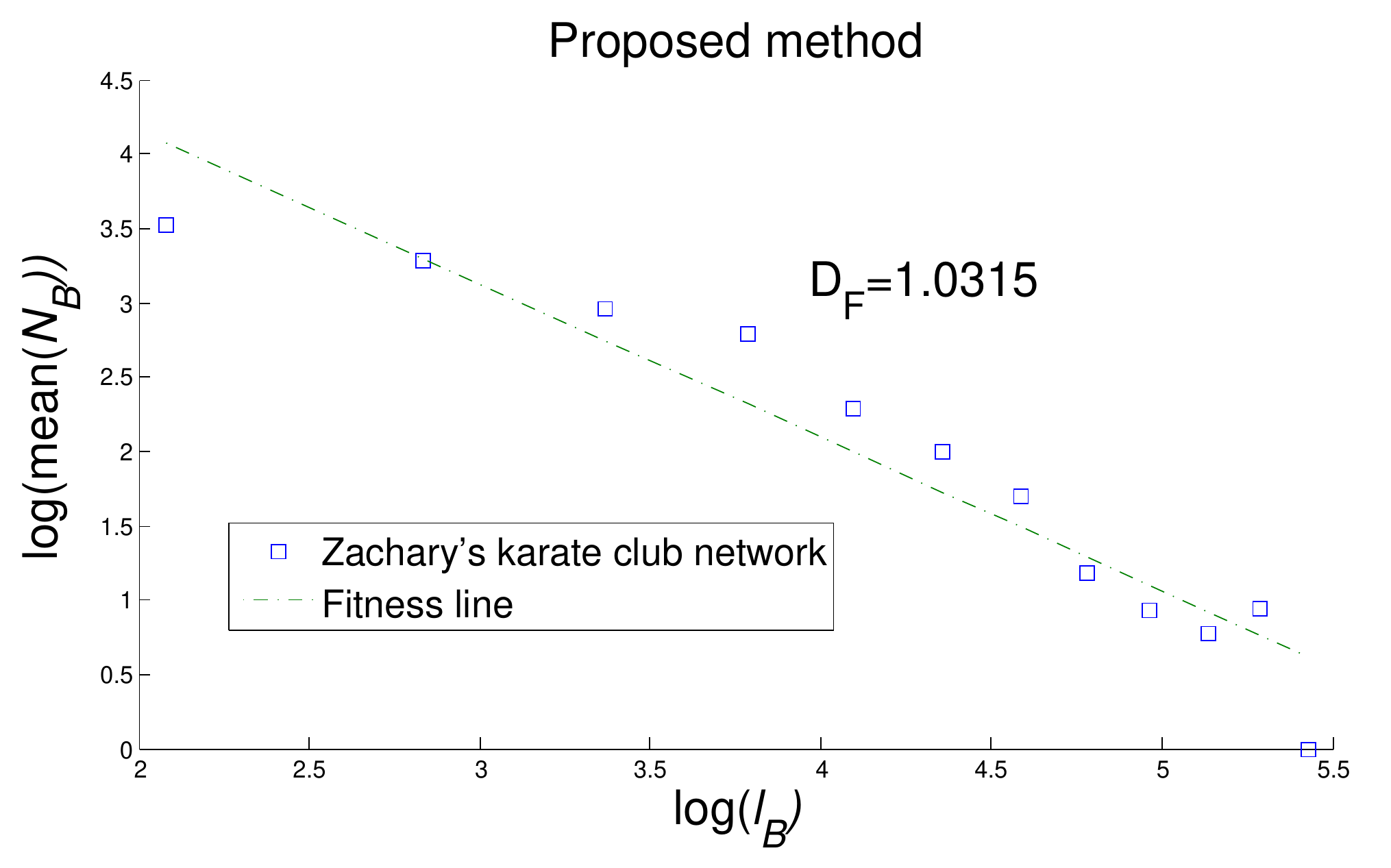}}
  \subfigure{
    \label{fig:subfig:b'} 
    \includegraphics[width=0.45\textwidth]{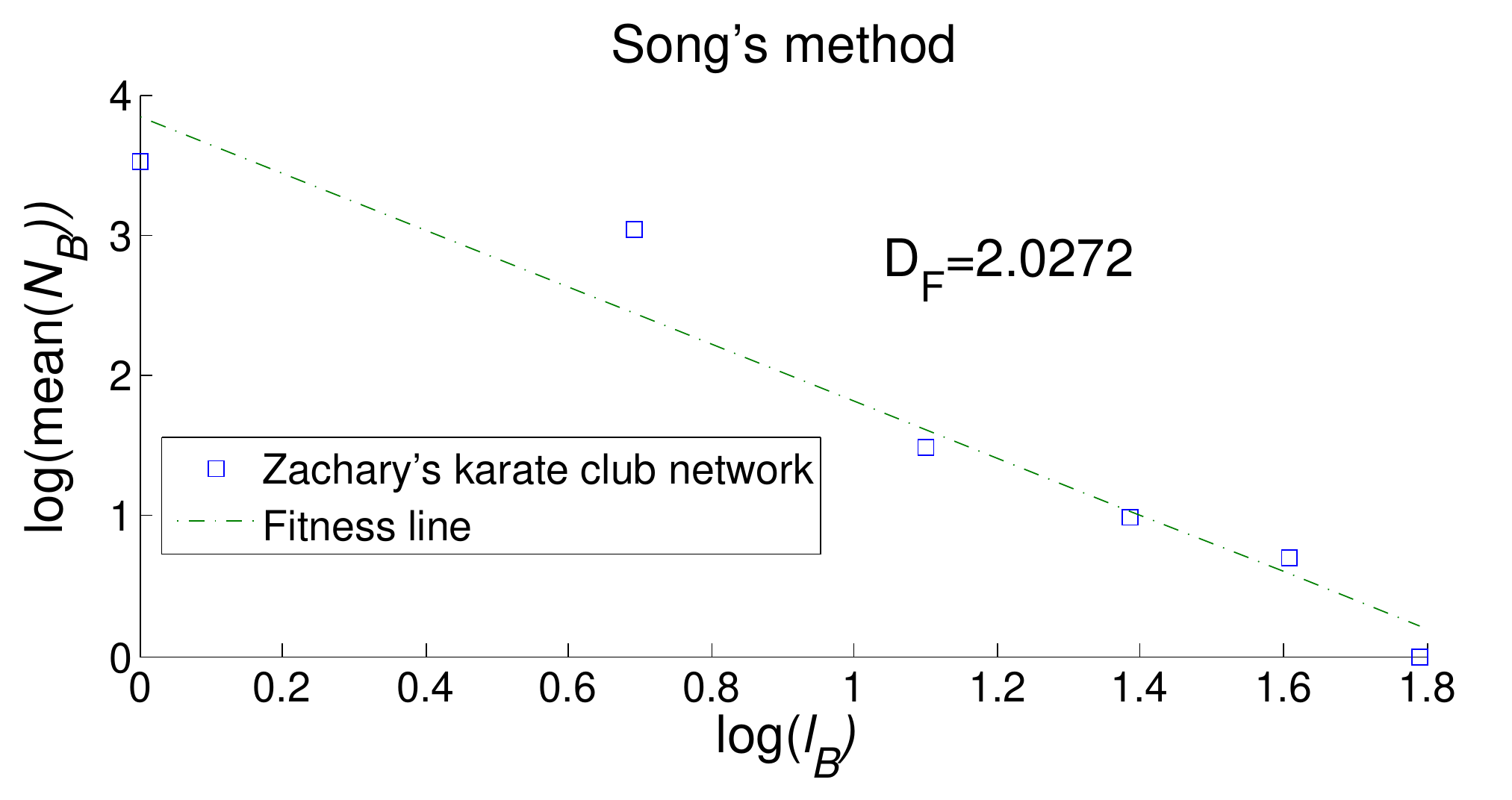}}
  \subfigure{
    \label{fig:subfig:c} 
    \includegraphics[width=0.45\textwidth]{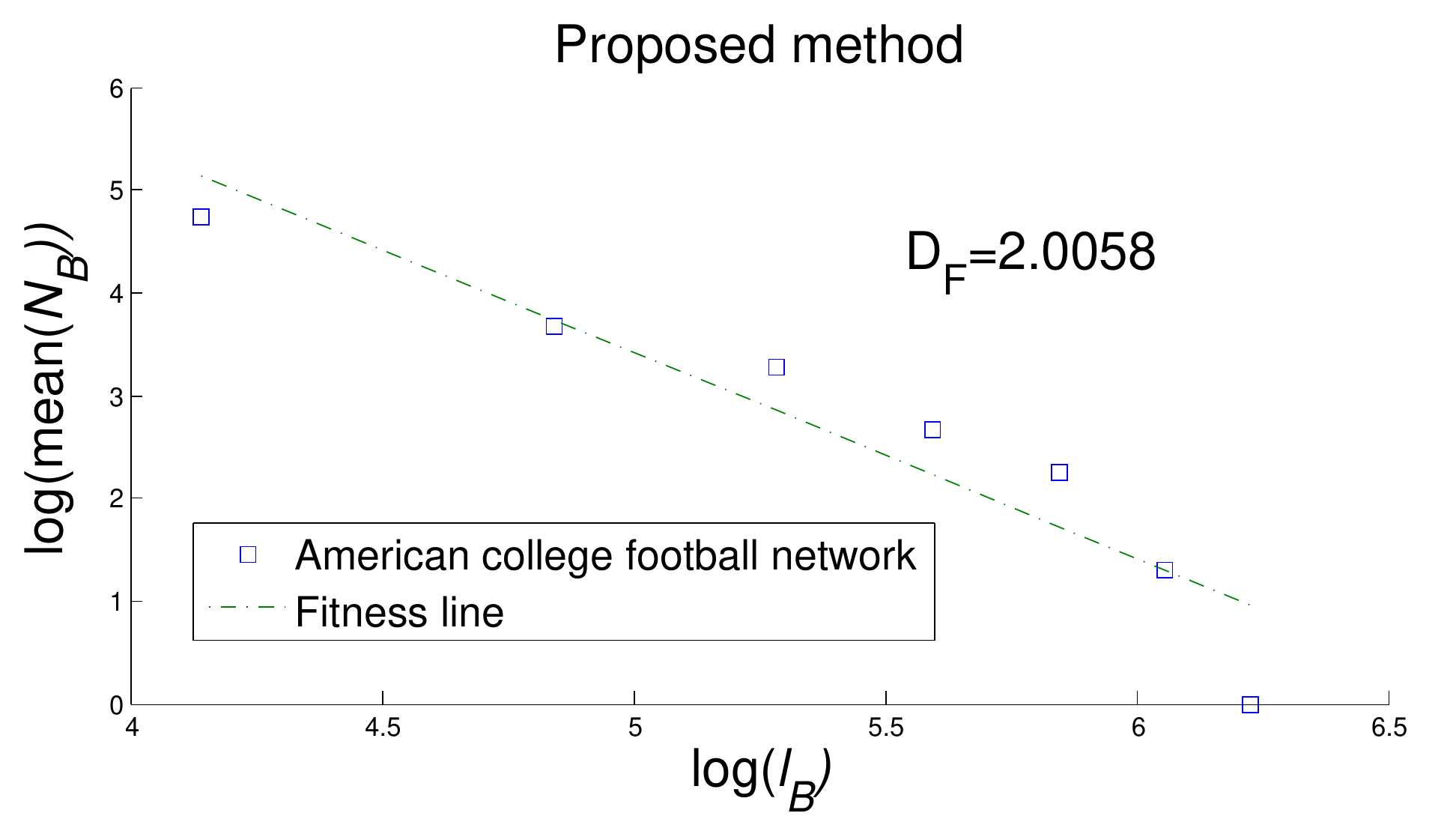}}
  \subfigure{
    \label{fig:subfig:c'} 
    \includegraphics[width=0.45\textwidth]{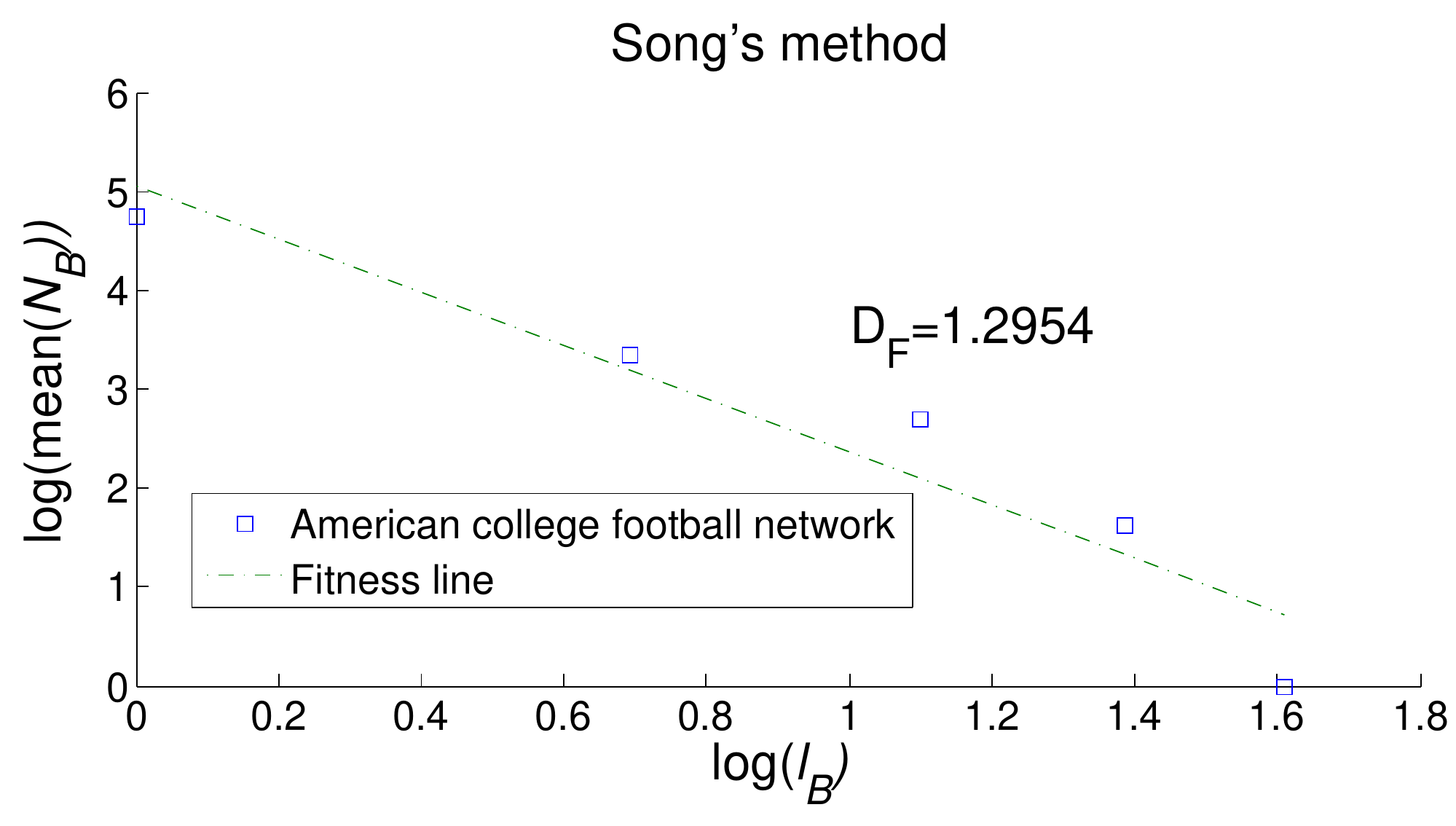}}
  \caption{The $N_{B}$ vs. $l_B$ of several real complex networks obtained in a log-log scale.The vertical ordinate of every subplot is the mean value of $N_{B}$ for 1000 times, and the horizontal ordinate represents the box size, $l_B$. Regressing $\log \left( {N_B } \right)$ vs. $\log \left( {l_B } \right)$ for different values of $l_B$, we obtained the fractal dimensions marked in every subplot both by our proposed method and Song's method.}
  \label{real complex networks} 
\end{figure}

In addition, several real complex networks are studied, they are the \emph{E.coli} network, with 2859 proteins and 6890 interactions between them\cite{schneider2012box}, Zachary's karate club, a social network of friendships between 34 members of a karate club at a US university in the 1970s\cite{zachary1977information}, American College football, a network of American football games between Division IA colleges during regular season Fall 2000\cite{girvan2002community}. The results are shown in Table \ref{table1} and Figure \ref{real complex networks}.

\begin{landscape}
\begin{table}
\begin{center}
  \caption{General characteristics of several real networks and the fractal dimensions $D_F$ obtained by our proposed method and Song's method. The fractal dimension is averaged for 1000 times, and the standard deviations of both methods are given in the parentheses for each network}
  \label{table1}
  \begin{tabular}{lllll}
  \toprule
        Network & Number & Number & Fractal dimension $D_F$ & Fractal dimension $D_F$ \\
        & of nodes & of edges & by our proposed method & by Song's method\cite{song2007calculate}\\
  \midrule
        E.coli network & 2859  & 6890  & 1.1361$\pm$0.0057905) & 3.4387($\pm$0.034737) \\
        American college football & 115   & 615   & 2.0087($\pm$0.035438) & 2.6941($\pm$0.038298) \\
        Zachary's karate club & 34    & 78    & 1.0364($\pm$0.029183) & 2.0294($\pm$0.025393) \\
  \bottomrule
\end{tabular}
\end{center}
\end{table}
\end{landscape}

%
From Figure \ref{real complex networks}, we can see that both our proposed method and Song's method could capture the self-similarity of these real complex networks. However, the standard deviations of the results of our proposed method are mostly less than that of Song's method. That is to say, our proposed method is more stable despite the influence of the original coloring sequence.

\section{Conclusions}
Hub repulsion is a key factor which leads to fractal complex networks. This paper models the self-similarity of complex network from the view of the hub repulsion. In our new model, the connected nodes in the complex networks are regarded as "electric charges" and there exists "electrostatic interaction force" between them over the edge. The connected nodes with higher degree have greater force over their link. A new box-covering algorithm is proposed to calculate the fractal dimension. The connected nodes with higher degree will be less likely to be covered by the same box in the box-covering process, which reflect the hub repulsion. The Sierpinski triangle network is studied. The results show that our new method is reasonable and efficient to caputure the self-similarity property of the Sierpinski triangle network. Some real complex networks were investigated, the results still shown the efficiency of our new model.

\section*{Acknowledgments}
We greatly appreciate the editor for the encouragement and the anonymous reviewers for their helpful comments and suggestions. The work is partially supported by National Natural Science Foundation of China, Grant Nos. 61174022 and 71271061, Chongqing Natural Science Foundation (for Distinguished Young Scholars), Grant No. CSCT, 2010BA2003, National High Technology Research and Development Program of China (863 Program) (No. 2012AA041101), Science and Technology Planning Project of Guangdong Province, China (2010B010600034, 2012B091100192), Business Intelligence Key Team of Guangdong University of Foreign Studies (TD1202), Doctor Funding of Southwest University Grant No. SWU110021.





\bibliographystyle{model1-num-names}
\bibliography{complexnetwork}







\end{document}